\documentclass{article}

\newcommand{\eq}{\begin{equation}}
\newcommand{\feq}{\end{equation}}
\newcommand{\eqn}{\begin{eqnarray}}
\newcommand{\feqn}{\end{eqnarray}}
\newcommand{\arr}{\begin{eqnarray*}}
\newcommand{\farr}{\end{eqnarray*}}
\newcommand{\beq}{\begin{equation}}
\newcommand{\eeq}{\end{equation}}
\newcommand{\bea}{\begin{eqnarray}}
\newcommand{\eea}{\end{eqnarray}}

\def\beq{\begin{equation}}
\def\eeq{\end{equation}}
\def\bea{\begin{eqnarray}}
\def\eea{\end{eqnarray}}
\def\bc{\begin{displaymath}}
\def\ec{\end{displaymath}}
\def\lb{\label}

\def\la{\lambda}

\def\lb{\label}

\begin{document}
\hbadness=10000

\begin{titlepage}

\begin{center}
\renewcommand{\thefootnote}{\fnsymbol{footnote}}
\vspace*{2.5cm}
{\Large \bf An Einstein-like theory of gravity with a non-newtonian 
weak-field limit}
\vskip 25mm
{\large \bf {M.~Cadoni \footnote{email: mariano.cadoni@ca.infn.it}}}\\
\renewcommand{\thefootnote}{\arabic{footnote}}
\setcounter{footnote}{0}
\vskip 7mm
{\small Universit\`a di Cagliari, Dipartimento di Fisica,\\
Cittadella Universitaria, 09042 Monserrato, Italy\\
and  INFN, Sezione di Cagliari}\\
\vspace*{0.4cm}
\end{center}
\vfill
\centerline{\bf Abstract}
\vskip 10mm
We propose a model describing Einstein gravity coupled 
to a scalar field with an exponential potential.
We show that the weak-field 
limit  of the model has  
static solutions given by a gravitational potential behaving for 
large distances  as $\ln r$. The 
Newtonian term $GM/r$ appears only as subleading.
Our model can be used to give a phenomenological explanation of the 
rotation curves of the galaxies without postulating the presence of 
dark matter. This can be achieved only by giving up at galactic scales 
Einstein  equivalence principle.   

\vskip 7mm
\end{titlepage}

The weak-field limit of Einstein general relativity is  Newton
theory of gravity. In every 
textbook on general relativity one learns that for weak and static 
gravitational fields in the non relativistic approximation, the 
Einstein equations become the Poisson Equation for the 
Newtonian potential $\bar\nabla^{2}\phi=4\pi G\rho$, which for a 
point-like source of mass $M$ has  
the solution $\phi= -MG/r$. 
Always in the same approximation, the geodesic equations of motion for a 
test particle in general relativity become Newton's second  law. 

The validity of the Newtonian limit of general relativity is 
unquestionable for distance scales ranging from the millimetre to 
solar system distances, where it is unambiguously supported by 
observations.  For larger distances the situation is more involved.
At galactic scales, the rotation curves of the galaxies cannot be explained 
by the Newtonian gravitational field generated by the visible matter 
(see for instance Ref. \cite{Trimble:ds}).
A Newtonian form of the potential is compatible with the observations 
only by postulating the existence of dark matter.
Conversely, the observed rotation curves of the galaxies can be 
explained, without postulating the existence of dark matter, by 
modifying the Newtonian dynamics at small accelerations \cite{Milgrom:1983ca}.
In particular, a non-newtonian gravitational potential behaving 
at galactic scales as $\phi\sim\ln r$ can explain the observational data.

More recently, radiometric data from the Pioneer, Galileo and Ulysses 
spacecrafts, have revealed anomalous accelerations, which could be 
explained by some modification of the Newtonian potential at small 
accelerations \cite{Anderson:1998jd}.

If one believes that the modification of Newtonian gravity at small 
acceleration is the right way to solve the puzzle of the rotation 
curves of the galaxies and the Pioneer anomaly, one has to find 
weak-field limits of general relativity (or some related theory of 
gravity) different from the Newtonian one.
This turns out to be a very complicated task.  It is easy to find 
Yukawa-like corrections to the Newtonian potential. This can be 
achieved for instance including higher-powers of the curvature tensor 
in the Einstein-Hilbert action \cite{Stelle:1977ry}.
To our knowledge, logarithmic corrections to the 
Newtonian potential  have been found only  in the context of a 
bimetric theory of gravity \cite{Bertolami:2003ui}. 
Unfortunately,   they appear  only as a subleading 
term of an asymptotically  linear  gravitational potential. 

In this paper 
we propose a model describing Einstein gravity coupled 
to a scalar field with an exponential potential.
We show that in the  weak-field 
limit our   model admits  
static solutions given by a gravitational potential behaving for large 
distances as $\ln r$.
The 
Newtonian term $-GM/r$ appears only as subleading.
Our model can be used to give a phenomenological explanation of the 
rotation curves of the galaxies without postulating the presence of 
dark matter. 
 Unfortunately, this can be achieved only by giving up at galactic 
 scales  a 
fundamental principle of the Einstein theory of gravity: the principle 
of equivalence.

We consider a system of two point-particles of mass $M$ and $m$, 
with $M>>m$ interacting with the gravitational field $g_{\mu\nu}$ and a scalar field $\varphi$.
The gravitational interaction is described by the Einstein action.
The scalar field has a potential $V(\varphi,\alpha)$, and its  
interaction with the point-particles is characterized by a coupling function 
$F(\varphi,\alpha)$.
Notice that both the potential $V$ and the coupling function $F$ 
depend not only on $\varphi$ but also on some real parameter $\alpha$.
Because $M$ is much bigger then $m$, 
the contribution of the mass $m$ as source of the gravitational field can be neglected.
Thus, the only sources   for 
the gravitational field are the mass $M$ and 
the scalar field $\varphi$.
The mass $m$ will be considered as a test particle, whose motion is 
determined by the field configuration for $g_{\mu\nu}$ and $\varphi$.

The system is described by the Einstein-like action (we use units, 
where the speed of light $c=1$ and a signature $(-1,1,1,1)$ for the 
metric) 

\bea\lb{action}
S&=&\int d^{4}x \sqrt{-g}\left[{1\over 16\pi G} R - \partial_{\nu}\varphi\partial^{\nu}\varphi
+V(\varphi,\alpha)\right]+\nonumber \\
&-&\sum _{a=1}^{2}m_{a}¥F_{a}¥(\varphi,\alpha)\int dt \sqrt{-g_{\mu\nu}{dx_{a}¥^{\mu}\over 
dt}{dx_{a}¥^{\nu}\over dt}},
\eea
where $m_{1}=M,\, m_{2}=m$ and $x_{a}(t)$ are the positions of the two 
point-particles.
The field equations describing the motion of the test particle with 
mass $m$
will be determined by taking variations of the action (\ref{action}) 
with respect to $g_{\mu\nu}$, $\varphi$
and $x_{2}$. We get

\bea\lb{fe}
&&R_{\mu\nu}-{1\over 2} g_{\mu\nu} R= 8\pi G\left[
T^{(\varphi)}_{\mu\nu}+(1+F)T^{(M)}_{\mu\nu}\right],\nonumber\\
&&2\nabla^{2}\varphi+ {\partial V\over\partial \varphi}=M{\partial F\over 
\partial\varphi} 
\int d\tau {\delta^{4}(x_{1}¥^{\nu}-x_{1}¥^{\nu}(\tau)\over\sqrt{-g}},\\
&&{d^{2}x_{2}¥^{\mu}\over d\tau^{2}¥}+\tilde\Gamma^{\mu}_{\rho\sigma}{dx_{2}¥^{\rho}\over d\tau}
{dx_{2}^{\sigma}\over d\tau}=0\nonumber.
\eea
where $T^{(\varphi)}_{\mu\nu}$,  $T^{(M)}_{\mu\nu}$ are the stress-energy tensors, 
respectively, for the scalar and for the source of mass $M$ and 
$\tilde\Gamma$ is a  $\varphi$-dependent connection. Notice 
that in  Eq. (\ref{fe}) we have neglected the contribution $T^{(m)}_{\mu\nu}$ 
of the test particle to the stress-energy tensor and to the equation for the scalar 
field and we have chosen $F_{2}=F_{1}-1=F$.
$T^{(\varphi)}_{\mu\nu}$ and  $T^{(M)}_{\mu\nu}$ are given by the following 
expressions
\bea\lb{Stress}
T^{(\varphi)}_{\mu\nu}&=& 2 
\partial_{\mu}\varphi\partial_{\nu}\varphi-g_{\mu\nu}\left[(\partial 
\varphi)^{2}-V\right],\\
T^{(M)}_{\mu\nu}&=&M\int d\tau 
u_{\mu}u_{\nu}¥{\delta^{4}(x_{1}¥^{\alpha}-x_{1}¥^{\alpha}(\tau))\over\sqrt{-g}},
\eea
where $u_{\mu}$ is the quadrivelocity of the particle.
$\tilde\Gamma$ is related to the usual affine connection
$\Gamma$ by the relation,
\beq
\tilde\Gamma^{\mu}_{\rho\sigma}=\Gamma^{\mu}_{\rho\sigma}+{1\over 
2F}\left(\partial _{\rho}F\delta^{\mu}_{\sigma}+\partial 
_{\sigma}F\delta^{\mu}_{\rho}-2g_{\rho\sigma}g^{\mu\gamma}\partial_{\gamma}F\right).
\feq

Let us now consider the usual weak-field, nonrelativistic, static 
limit of the field equations (\ref{fe}). Setting $g_{\mu\nu}= 
\eta_{\mu\nu}+h_{\mu\nu}$ with $h_{\mu\nu}<<1$, considering field configurations 
depending only on the spatial coordinates $ x^{i},\, 
i=1,2,3$, in the nonrelativistic limit, when the velocity of the 
particles $ v<<1$ and $|T_{ij}|<<|T_{00}|$, the field equations (\ref{fe}) give,
\bea\lb{nl}
\bar\nabla^{2}\psi&=&4\pi G\left\{\left[(\bar\nabla 
\varphi)^{2}-V\right]+ (1+F) \tilde T^{(M)}_{00}¥\right\}\nonumber\\
\bar\nabla^{2}\varphi&=&{1\over 2}\left( {
\partial F\over \partial \varphi} \tilde T^{(M)}_{00}- {
\partial V\over \partial \varphi}\right)\\
{d^{2}\bar x\over dt^{2}}&=&-\bar\nabla\psi -
{1\over F}{\partial F\over \partial \varphi} \bar\nabla\varphi\nonumber,
\eea
where $\psi=-h_{00}/2$, $\tilde T^{(M)}_{00}= M\delta^{3}(\bar x- \bar
x_{1})$, the bar indicates three-dimensional vectorial quantities, we 
have set $\bar x_{2}=\bar x$ 
and   the differential operators are calculated with respect to the 
three-dimensional Euclidean metric.

The usual weak-field Newtonian limit can be trivially recovered setting 
in Eqs. (\ref{nl}) $F=0,V=0$ and picking the $\varphi=0$ solution for 
the scalar field equation.  
It is important to notice that  there is an other way to recover the 
Newtonian limit from Eqs (\ref{nl}). 
Setting $F=0$ and choosing a potential $V$, which allows for solutions satisfying   
$(\bar\nabla¥ 
\varphi)^{2}¥=V$, the scalar field  decouples from the gravitational 
sector. The first and the third equations in (\ref{nl}) become, 
respectively, the Poisson Equation and Newton's second law.
When the potential satisfies  the equation $(\bar\nabla¥ 
\varphi)^{2}¥=V$ consistency with the relation  $|T_{ij}|<<|T_{00}|$, 
which determines the nonrelativistic  limit, requires that the scalar 
field changes very slowly on the scale of distances we are 
considering. Using for instance spherical coordinates this means that
the term $(\partial_{r}\varphi)^{2}$ in the $rr$ component of the 
stress-energy tensor can 
be neglected.

Let us now choose  a potential and a coupling function with an exponential 
form
\beq\lb{e1}
V=\lambda^{2}¥\exp{\left(-{4\sqrt\pi\over \alpha}\varphi\right)},\quad 
F=\exp{\left(2 \sqrt{\pi} \alpha G\varphi\right)},
\eeq
where $\lambda^{2}¥$ is a constant with dimensions 
(mass)(length)$^{-3}$. Because in the action (\ref{action}) we define   
the potential $V$ with a sign opposite to the standard definition, 
our choice of Eq. (\ref{e1}) corresponds to a negative potential.
A model  of Einstein gravity coupled to scalar field with a 
negative  exponential potential
has been already  proposed  in the literature for solving the
problem of the rotation curves of the galaxies \cite{Matos:2000ki, Matos:2001pz}.
The main difference between our model and that considered in Ref. \cite{ Matos:2000ki})
is the fact that we  introduce the coupling function 
$F(\varphi,\alpha)$.

Using Eqs. (\ref{e1}) and defining the new 
field
\beq\lb{e2}
\phi= \psi+2\sqrt{\pi} G\alpha \varphi.
\feq
Eqs. (\ref{nl}) become
\bea\lb{nl1}
\bar\nabla^{2}\phi&=&4\pi G\left\{(\bar\nabla 
\varphi)^{2}+ \left[1+(1+ G \alpha^{2})e^{2\sqrt{\pi} G\alpha \varphi}¥\right] \tilde T^{(M)}_{00}¥
\right\},\nonumber\\
\bar\nabla^{2}\varphi&= &2\sqrt\pi\left( G \alpha e^{2\sqrt \pi G\alpha \varphi} \tilde T^{(M)}_{00}
+ {\lambda^{2}¥\over \alpha} e^{-{4\sqrt\pi\over 
\alpha}\varphi}\right),\\
{d^{2}\bar x\over dt^{2}}&=&-\bar\nabla\phi\nonumber .
\eea

From the third  Equation in  (\ref{nl1}) it is evident that the field $\phi$ represents the 
potential that determines the force acting on the test particle.  A 
spherical symmetric 
solution  to the Eqs. (\ref{nl1}) can be found placing the 
source-particle of mass $M$ at the origin of the coordinate system and 
using spherical coordinates $(r,\theta, \omega)$. The solution reads
\beq\lb{sol}
\varphi={\alpha\over 2\sqrt{\pi}} \ln\left({2\sqrt{\pi}\lambda\over \alpha} r\right),
\feq\
\beq\lb{sol1}
\phi= G\alpha^{2} \ln Cr- {GM\over r},
\eeq
where $C$ is an arbitrary integration constant.
The  test particle will experience an acceleration
\beq\lb{sol3}
 a= -{d\phi\over dr}=- { G\alpha ^{2}\over r} -{G M\over 
r^{2}}.
\feq
The potential $\phi$ of Eq. (\ref{sol1}) has the $-GM/r$ Newtonian behavior 
only near $r=0$. Its asymptotical, $r\to\infty$,  behavior is 
logarithmic and therefore  
radically non-newtonian. Far away from the source the Newtonian term is only subleading.  
The unpleasant feature  of the gravitational potential 
(\ref{sol1}) is its dependence from the parameter $\alpha$  
parametrizing both the potential $V$ for the scalar field and the 
coupling function $F$. Our model can be phenomenologically relevant 
only if  
the logarithmic term appearing in
Eq. (\ref{sol1})  depends on the mass $M$ of the source.
Moreover, to preserve the standard results of general relativity and 
its Newtonian limit at solar system scales, $\alpha$ should depend 
also on some threshold acceleration $a_{0}$, whose magnitude is such 
that the logarithmic term in Eq. (\ref{sol1}) becomes relevant only 
at galactic scales.
Formally, this can be achieved by writing the parameter $\alpha$ as 
function of $M$ and of the  
constants  $G, \la$: $\alpha=\alpha(M,G,\lambda)$.
If this is the case our model (\ref{action}) can be used to solve the 
problem of the rotation curves of the galaxies without postulating the 
presence of dark matter.

The rotation curves at distance $r$ from the 
galactic core can be described by the equation
\beq\lb{en2}
v^{2}(r)= {GM(r)\over r},
\feq
where $v(r)$ is the velocity of a layer at distance $r$ and $M(r)$ is 
the total mass inside the layer.
Observations are consistent with $M(r)$ behaving 
as 
\beq\lb{en3} 
M(r)= A r + B,
\feq
where $A$ and $B$ are some constants (see for instance Ref.\cite{Trimble:ds}). 
Using equations (\ref{sol3}) and Eq. (\ref{en2}) one easily derives 
\beq\lb{ef4}
M(r)=  \alpha^{2} r + M,
\feq
in accordance with the experimental curve (\ref{en3}).

Fixing appropriately the form of the constant $\alpha$ our model can 
be used to derive the modified Newtonian dynamics (MOND) of Milgrom 
 \cite{Milgrom:1983ca} 
as the weak-field limit of the Einstein-like model (\ref{action}).
MOND  introduces a constant acceleration $a_{0}\sim 10^{-29}cm^{-1}¥ $, 
such that the 
standard Newtonian dynamics is a good approximation only for 
accelerations $a>>a_{0}$. 
For $a \sim  a_{0}$ MOND  predicts  that a test particle at 
distance $r$ from a mass $M$ experiences an acceleration \cite{Milgrom:1992hr}
\beq\lb{mond}
|a|={\sqrt{Ma_{0}¥G}\over r}.
\feq

This expression can be derived from our model setting  
\beq\lb{en4}
\alpha^{2}= \sqrt{Ma_{0}\over G}.
\feq
Using this equation we can  see that the leading 
term  in Eq. (\ref{sol3}) reproduces exactly the MOND result 
(\ref{mond}). Simple dimensional analysis allows us to
identify $a_{0}$ in terms of the two dimensional parameters 
$G,\lambda$ appearing in our model : $a_{0}=\lambda 
\sqrt{G}$.

For $a>>a_{0}$ we recover the standard Newtonian dynamics.
In fact, $a>>a_{0}$ implies 
$\sqrt{GM}/r>>\sqrt{a_{0}}$. It follows that in this limit in Eq. 
(\ref{sol3}) the Newtonian term $GM/r^{2 }$ dominates with respect to the first 
term.   Eq. (\ref{sol}) tells us that for   $a>>a_{0}$, 
$\varphi\to -\infty$, which  in turn implies $F=0$. Because the 
solution (\ref{sol}) satisfies $(\bar \nabla \varphi)^{2}¥=V$, the scalar 
field decouples in the weak-field limit from the gravitational sector 
and we obtain the standard Newtonian limit (see discussion after 
Eq. (\ref{nl})).

The Einstein theory of gravity with a negative cosmological constant 
can be obtained as a particular case of our model.
Taking in Eqs. (\ref{e1}) the limit $\alpha\to -\infty$ we have $F=0$ and $V=\lambda^{2}$.
Picking the $\varphi=0$ solution of the field equation for the scalar, 
the action (\ref{action}) becomes the Einstein-Hilbert action with a 
cosmological constant.

It is obvious that identifying the parameter $\alpha$ in terms of 
the mass of the source we are giving up at galactic scales  a fundamental principle of the 
Einstein theory of gravity: the principle of equivalence. This is not 
immediately evident in our simplified model (\ref{action}) because we
are  considering the motion of a test particle in the gravitational field generated 
by the source of mass $M$. The source is completely characterized by
its   gravitational mass whereas for the test particle only its inertial mass
can be relevant. Because the equations of motion of the test 
particle (\ref{fe}) turn out to
be independent of its mass, one could be erroneously led to conclude 
that the equivalence principle still holds. 
This is  not true. The breakdown of the equivalence principle 
will immediately show up when we  try to describe in a self consistent 
way the mutual interaction of the two masses.

Apart from the breakdown of the equivalence principle, a gravitational theory 
described by the action (\ref{action}) in which 
the parameter $\alpha$ is a function of the mass $M$ of the source  
poses also huge interpretation problems. 
Implicitly we are assuming the existence of a ``cosmic'' scalar field 
$\varphi$ whose self-coupling (the potential $V$) and its coupling with 
the matter (the coupling function $F$) are determined  by the 
distributions of the sources for the gravitational field.
We  do not have a definite prescription of how the information about 
the distribution of matter has to be encoded on the form of the 
functions $F$ and $V$. There is no general argument, no principle,
behind our Eq. (\ref{en4}). Its only justification is the accordance 
with the observed rotation curves of the galaxies. 
For this reason our model, at least in the present context, cannot 
have a fundamental but just a phenomenological character.
Independently of the fundamental, 
still unknown  physics that could lie behind our phenomenological 
model, it is likely  that the information about the distribution of 
matter has to be encoded
in the cosmic field,  trough the form of the functions  
$V(\varphi,M)$ and $F(\varphi,M)$,  in a non local way.
We are leaving the Einsteinian paradigm and moving toward a Machian 
description of the gravitational interaction.

Our  model (\ref{action}) can be considered as a particular case of 
a scalar-tensor theory of gravity.
It is well known that scalar-tensor theories of 
gravity can reproduce the standard phenomenology of general relativity at 
solar-system scales (perihelion shift of Mercury and bending of light 
by the sun), only for particular values of the parameters entering in 
the theory \cite{Wagoner:1970vr}. It is therefore necessary to check that our model 
(\ref{action}) with  potential  and coupling function 
given by Eq. (\ref{e1}), apart from 
explaining the rotational curves of the galaxies,  can also  get 
through  the standard tests of general relativity at solar-system scale.
This point will be discussed in a forthcoming publication.

We conclude by noticing the striking similarities of our model with 
the ``quintessence'' models proposed in cosmology for solving the 
dark energy problem \cite{Caldwell:1997ii, Barreiro:1999zs}.  
This problem can be solved 
by introducing in the 
Einstein theory a scalar field (the quintessence field) with an 
exponential potential. The main difference with our model, apart from 
the presence of the coupling function $F$, is the sign 
of the potential, which using the standard notation is positive for quintessence 
models and 
negative  for our model.  
It is amusing that both  dark matter and 
dark energy problem can be solved, at least phenomenologically, by 
introducing in the Einstein action scalar fields with exponential 
potentials.

\begin{flushleft}

{\bf Acknowledgments}
\end{flushleft}
We thank S. Mignemi, M. Cavagli\'a,  M. Lissia and D. L. Wiltshire for discussions 
and valuable comments.

\end{document}